\documentclass[prl,10pt,a4paper,nosuperscriptaddress,twocolumn,noshowpacs,notitlepage,noshowkeys]{revtex4}

\usepackage{times}
\usepackage{amsmath}
\usepackage{amssymb}
\usepackage{graphicx}

\newcommand{\be}{\begin{equation}}
\newcommand{\ee}{\end{equation}}
\newcommand{\ben}{\begin{eqnarray}}
\newcommand{\een}{\end{eqnarray}}
\newcommand{\bes}{\begin{subequations}}
\newcommand{\ees}{\end{subequations}}

\begin{document}
\title{Entropy of entangled three-level atoms interacting with entangled cavity fields: \\entanglement swapping}
\author{Wen-Chao Qiang$^{a,}$\footnote{Corresponding author.\\E-mail address: qwcqj@pub.xaonline.com (W.-C. Qiang).}, W.B. Cardoso$^b$, Xin-Hui Zhang$^a$}
\affiliation{$^a$Faculty of Science, Xi'an University of
Architecture and Technology, Xi'an, 710055, China\\
$^b$Instituto de F\'{\i}sica, Universidade Federal de Goi\'{a}s, 74.001-970, Goi\^{a}nia,Goi\'{a}s, Brazil}

\pacs{89.70.Cf, 03.67.Bg, 42.50.Ex, 03.65.Ud}
\keywords{Three-level atom, Jaynes-Cummings model, Full microscopic
Hamiltonian approach, Entropy}

\begin{abstract}
The dynamics of an entangled  atomic system in a partial interaction with entangled cavity fields, characterizing an entanglement swapping, have been studied through the use of Von Neuman entropy. We consider the interaction via two-photon process given by a full microscopical Hamiltonian approach. The explicit expression of the entropy  is obtained, wherewith we estimated the largest period. The numerical simulation of the entropy of the entangled atomic and cavity systems shows that its time evolution presents multi-periodicity. The effects of detuning parameter on the period and the amplitude of the  entropy are also discussed.
\end{abstract}

\maketitle

\section{Introduction}
Like the Shannon entropy measures the uncertainty associated with a classical probability distribution,  the Von Neuman entropy describes a quantum state via density operators [1].
It is a very useful  measure of the purity of quantum states and contains all moments of the density operator. Therefore, the entropy can be used as a measure of the entanglement degree of quantum states  and has been extensively used to study the interaction between light field and atoms via Jaynes-Cummings model [2,3]. On the other hand, even though the entropy given an estimate of the entanglement degree, but it is not sufficient, such as the negativity or the concurrence. The problem of the last two is that they only apply to 2x2 or 2x3 systems, which is not our case. The entropy of the Jaynes-Cummings system has attracted much attention [4-9]. The expression of the field entropy for the entangled states of a single two-level atom interacting with a single electromagnetic field mode in an ideal cavity with the atom undergoing either a one or a two-photon transition has been obtained in [9]. In addition, the evolution of the atomic or field entropy for a three-level atom interacting with one-mode [10,11] and two-mode [12-15] of cavity fields has also been examined. Furthermore, recently Obada and Eied studied the entropy of a system composed of a $\Xi$-type three-level atom interacting with a non-correlated two-mode cavity field in the presence of nonlinearities [16].

Though all works mentioned above revealed a lot of interesting properties of the Jaynes-Cummings system, they limited to the situation of a single atom interacting with a cavity field. The study of the entropy of double-atom interacting via two-photon process with single or double cavity fields is missing to our knowledge. In a very recent work, dSouza et al studied the entanglement swapping in the two-photon Jaynes-Cummings model [17]. Preparing atoms 1 and 2 in an entangled state and two cavity fields 3 and 4 also in other entangled state, then sending atom 2 through cavity 3 during a proper interaction time, they implemented the entanglement
swapping of the atomic and cavity systems. Motivated by this result, in the present work, we  will study  the entropy of the field interacting with two three-level atoms during the entanglement swapping protocol.

This paper is organized as follows. We give a brief introduction to our Hamiltonian model and its solution in the next section. The key part, section 3 devotes to find analytical  expression of the entropy of the atomic  and cavity systems. Some numerical simulations and discussions are given in section 4. We finally present our conclusions in section 5.

\section{Overview of the model}
In this paper, we use the so-called `full microscopic
Hamiltonian approach' (FMHA) [18,19] to describe a three-level atom in
$\Xi$ -configuration interacting with a single mode of a cavity-field via two-photon Jaynes-Cummings model.  In the absence of a driven field upon the atom, the Hamiltonian (interaction picture) that describes the atom-field interaction is given by
\begin{eqnarray}\label{1}
H_{I} &=&\hbar g_{1}\left( a|e\rangle \langle f|e^{-i\delta
t}+a^{\dagger}
|f\rangle \langle e|e^{i\delta t}\right)   \nonumber \\
&&+\hbar g_{2}\left( a|f\rangle \langle g|e^{i\delta t}+a^{\dagger
}|g\rangle \langle f|e^{-i\delta t}\right),
\end{eqnarray}
where $g_1$ and $g_2$ stand for the one-photon coupling constant
with respect to the transitions $|e\rangle \leftrightarrow |f\rangle$ and $|f\rangle \leftrightarrow |g\rangle$ respectively. $\delta$ is detuning parameter defined as:
\begin{equation} \label{2}
\delta =\Omega -(\omega _{e}-\omega _{f})=(\omega _{f}-\omega
_{g})-\Omega,
\end{equation}
where $\Omega$ denotes the cavity-field frequency and $\omega_e, \omega_f$ as well as $\omega_g$ are the frequencies associated with the atomic levels $|e\rangle, |f\rangle$ and
$|g\rangle$, respectively. Fig.1 is a schematic representation
of the atomic levels.

\begin{figure}
\centering
\includegraphics[width=3cm]{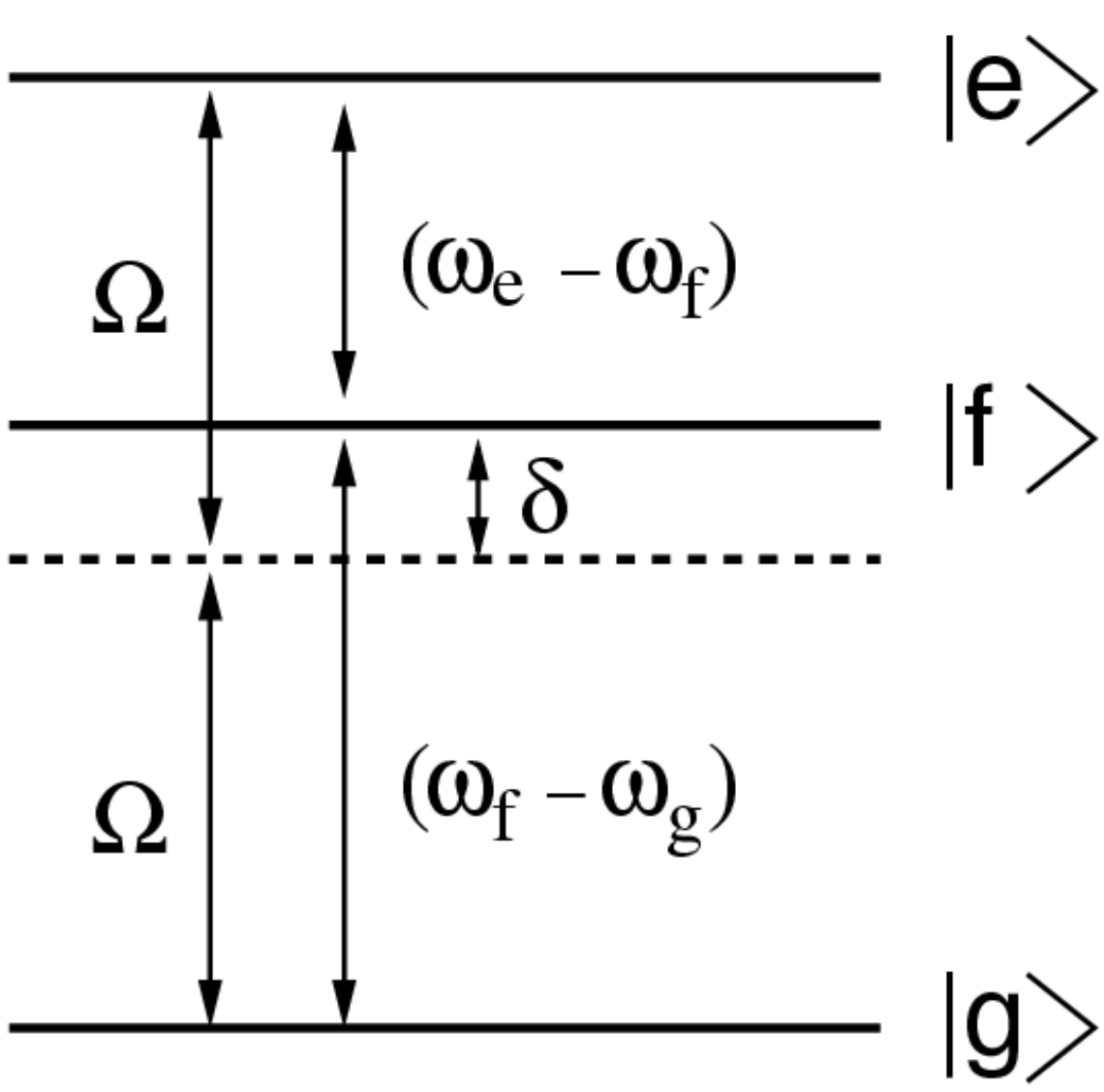}
\caption{Schematic diagram of the level structures of a three-level atom interacting with a single-mode of a cavity field.} \label{levels}
\end{figure}

The state of the composed atom-field system can be written as
\begin{equation}\label{3}
|\psi (t)\rangle =\sum_{n}\left[ C_{e,n}(t)|e,n\rangle
+C_{f,n}(t)|f,n\rangle +C_{g,n}(t)|g,n\rangle \right] ,
\end{equation}

Solving  the time dependent Schr\"{o}dinger equation with
Eqs.(\ref{1}) and (\ref{3}) we obtain the probability
amplitudes
\begin{eqnarray}\label{4}
C_{e,n}(t)&=&\left[ \frac{g_{1}^{2}(n+1)}{\Lambda _{n}\alpha
_{n}^{2}}\gamma _{n}(t)+1\right] C_{e}C_{n} \nonumber \\
&&-i\frac{g_{1}\sqrt{n+1}}{\Lambda _{n}}\sin
(\Lambda _{n}t)e^{-i\frac{\delta}{2}t}C_{f}C_{n+1} \nonumber \\
&&+\left[ \frac{g_{1}g_{2}
\sqrt{(n+1)(n+2)}}{\Lambda _{n}\alpha _{n}^{2}}\gamma
_{n}(t)\right] C_{g}C_{n+2},
\end{eqnarray}

\begin{eqnarray}\label{5}
C_{f,n+1}(t)&=&-i\frac{g_{1}\sqrt{n+1}}{\Lambda _{n}}\sin (\Lambda _{n}t)e^{i%
\frac{\delta}{2}t}C_{e}C_{n}\nonumber \\
&&+\left( \cos (\Lambda _{n}t)-\frac{i\delta }{
2\Lambda _{n}}\sin (\Lambda _{n}t)\right) e^{i\frac{\delta}{2}t}C_{f}C_{n+1} \nonumber \\
&&-i\frac{g_{2}\sqrt{n+2}}{\Lambda _{n}}\sin (\Lambda _{n}t)e^{i\frac{\delta}{2}t}C_{g}C_{n+2},
\end{eqnarray}

\begin{eqnarray}\label{6}
C_{g,n+2}(t)&=&\frac{g_{1}g_{2}\sqrt{(n+1)(n+2)}}{\Lambda _{n}\alpha _{n}^{2}}%
\gamma _{n}(t)C_{e}C_{n}\nonumber\\
& & -i\frac{g_{2}\sqrt{n+2}}{\Lambda_{n}}\sin (\Lambda
_{n}t)e^{-i\frac{\delta}{2}t}C_{f}C_{n+1} \nonumber \\
&&+\left[ \frac{g_{2}^{2}(n+2)}{\Lambda _{n}\alpha _{n}^{2}}\gamma _{n}(t)+1\right] C_{g}C_{n+2},
\end{eqnarray}
where
\begin{equation}\label{7}
\gamma _{n}(t)=\left[ \Lambda _{n}\cos (\Lambda _{n}t)+i\frac{\delta }{2}
\sin (\Lambda _{n}t)-\Lambda _{n}e^{i\frac{\delta }{2}t}\right] e^{-i\frac{
\delta}{2}t},
\end{equation}
\begin{equation}\label{8}
\Lambda _{n}=\sqrt{\frac{\delta ^2}{4}+\alpha_{n}^2},
\end{equation}

\begin{equation}\label{9}
\alpha _{n}=\sqrt{g_{1}^{2}(n+1)+g_{2}^{2}(n+2)},
\end{equation}
$\Lambda _{n}$ is  the Rabi frequency. $C_n\equiv C_{n}(0)$
stand for the amplitudes of the arbitrary initial field state and
the $C_{a}(a=e,f,g)$ are atomic amplitudes of the (normalized)
initial atomic state

\begin{equation}\label{10}
|\chi \rangle =C_{e}|e\rangle +C_{f}|f\rangle +C_{g}|g\rangle.
\end{equation}
Now, let us suppose two atoms (1 and 2) and two cavities (3 and 4) initially  in the states respectively
\begin{equation}\label{11}
    |\phi\rangle_{12}=\alpha_1|g\rangle_1|e\rangle_2+\beta_1|e\rangle_1|g\rangle_2,
\end{equation}
\begin{equation}\label{12}
    |\psi\rangle_{34}=\alpha_2|2\rangle_3|0\rangle_4+\beta_2|0\rangle_3|2\rangle_4.
\end{equation}
where $|0\rangle$ and $|2\rangle$ indicate the vacuum and two-photon states in the cavities, respectively.
Then, the atom 2 is sent to the cavity 3 during an interacting time $t$. Finally the total system state develops into
\begin{eqnarray}\label{13}
  \nonumber   |\Psi\rangle_{1234} &=& \alpha_1\alpha_2 |g\rangle_1|0\rangle_4\left [C_{e,2}^{(e,2)}(t)|e\rangle_2|2\rangle_3 \right.\\
  \nonumber    &&+\left. C_{f,3}^{(e,2)}(t)|f\rangle_2|3\rangle_3 +C_{g,4}^{(e,2)}(t)|g\rangle_2|4\rangle_3\right ]\\
\nonumber      &&+\alpha_1\beta_2 |g\rangle_1|2\rangle_4\left [C_{e,0}^{(e,0)}(t)|e\rangle_2|0\rangle_3 \right. \\
\nonumber      && +\left. C_{f,1}^{(e,0)}(t)|f\rangle_2|1\rangle_3 +C_{g,2}^{(e,0)}(t)|g\rangle_2|2\rangle_3\right ] \\
\nonumber      && +\alpha_2\beta_1 |e\rangle_1|0\rangle_4\left [C_{g,2}^{(g,2)}(t)|g\rangle_2|2\rangle_3 \right.\\
\nonumber      && +\left. C_{f,1}^{(g,2)}(t)|f\rangle_2|1\rangle_3 +C_{e,0}^{(g,2)}(t)|e\rangle_2|0\rangle_3\right ] \\
               && \beta_1\beta_2|e\rangle_1|2\rangle_4|g\rangle_1|0\rangle_3,
\end{eqnarray}
where $C_{x,n}^{(y,m)} ((x,y=e,f,g)$ and $(n,m=0,1,2))$ denotes the amplitude probability given by equation (4-6) considering the initial state in (y,m) with $x$ and $y$ representing the atomic state, but $n$ and $m$ the field state. For example, from equation (4) and (5) we have
\begin{equation}\label{14}
    C_{e,2}^{(e,2)}(t) = 1+\frac{3 g_1^2 \gamma_2(t)}{\alpha_2^2 \Lambda_2},
\end{equation}
\begin{equation}\label{l5}
    C_{f,1}^{(g,2)}(t) = -i \frac{\sqrt{2} g_2 \sin(\Lambda_0 t)}{\Lambda_0}e^{i \delta t/2}.
\end{equation}
In the next section, we shall study the entropy of the system in this state.

\section{Entropy}

After  the entanglement swapping, the density operator of the whole system is written in the form
\begin{equation}\label{16}
   \rho= |\Psi\rangle_{1234}~ {}_{1234}\langle\Psi|.
\end{equation}
It is easy to see from Eqs.(11) and (12) that both atomic and cavity field systems are in pure states at initial moment and  independent of each other. Therefore, the entropy of the whole system of field and atom is zero and remains constant. On the other hand, according to Araki-Lieb inequality [20], the entropies of the atomic system and the cavity system satisfy
\begin{equation}\label{17}
   |S_a-S_f|\preceq S\preceq |S_a+S_f|.
\end{equation}
Since $S=0$ at any moment $t>0$, entropy of the cavity system equals that of the  atomic system. So, in the following we only need to calculate the entropy of the  atomic system. This can be easily done. Taking partial trace of $\rho$ with cavity 3 and 4, we obtain the reduced density operator of the two atom system
\begin{widetext}
\begin{eqnarray}\label{18}
   \rho_{12}&=&|\alpha_2 \beta_1 C_{e,0}^{(g,2)}(t)|^2|e,e\rangle_{12}\cdot {}_{12}\langle e,e| +|\alpha_2 \beta_1 C_{f,1}^{(g,2)}(t)|^2|e,f\rangle_{12}\cdot {}_{12}\langle e,f| +|\beta_1|^2 [|\beta_2|^2 + |\alpha_2 C_{g,2}^{(g,2)}(t)|^2]|e,g\rangle_{12}\cdot {}_{12}\langle e,g|\nonumber\\
            &+&\alpha_1\beta_1^* [|\beta_2|^2 C_{e,0}^{(e,0)}(t) +|\alpha_2|^2 C_{e,2}^{(e,2)}(t)C_{g,2}^{(g,2)}(t)^*]|g,e\rangle_{12}\cdot {}_{12}\langle e,g|+\beta_1\alpha_1^*[|\beta_2|^2 C_{e,0}^{(e,0)}(t)^*+|\alpha_2|^2 C_{g,2}^{(g,2)}(t) C_{e,2}^{(e,2)}(t)^*]\nonumber\\
            &\times & |e,g\rangle_{12}\cdot {}_{12}\langle g,e| +|\alpha_1|^2 [|\beta_2 C_{e,0}^{(e,0)}(t)|^2 +|\alpha_2 C_{e,2}^{(e,2)}(t)|^2] |g,e\rangle_{12}\cdot {}_{12}\langle g,e| +|\alpha_1|^2 [|\beta_2 C_{f,1}^{(e,0)}(t)|^2 +|\alpha_2 C_{f,3}^{(e,2)}(t)|^2] \nonumber\\
            &\times & |g,f\rangle_{12}\cdot {}_{12}\langle g,f| +|\alpha_1|^2 [|\beta_2 C_{g,2}^{(e,0)}(t)|^2 +|\alpha_2 C_{g,4}^{(e,2)}(t)|^2] |g,g\rangle_{12}\cdot {}_{12}\langle g,g|
\end{eqnarray}
\end{widetext}
Using orthogonal basis $|a,b\rangle=|a\rangle_1|b\rangle_2 (a,b=e,f,g)$, we obtain a $9\times9$ matrix expression of $\rho_{12}$ with following non-zero elements:
\begin{eqnarray}\label{19}
\rho_{gg,gg}&=& |\alpha_1|^2 [|\beta_2 C_{g,2}^{(e,0)}(t)|^2
   + |\alpha_2 C_{g,4}^{(e,2)}(t)|^2],\nonumber\\
\rho_{gf,gf} &=& |\alpha_1|^2 [|\beta_2 C_{f,1}^{(e,0)}(t)|^2
   + |\alpha_2 C_{f,3}^{(e,2)}(t)|^2], \nonumber\\
\rho_{ge,ge} &=& |\alpha_1|^2 [|\beta_2 C_{e,0}^{(e,0)}(t)|^2
   + |\alpha_2 C_{e,2}^{(e,2)}(t)|^2],  \nonumber\\
\rho_{ge,eg} &=& \alpha_1 \beta_1^* [|\beta_2|^2 C_{e,0}^{(e,0)}(t)+|\alpha_2|^2 C_{e,2}^{(e,2)}(t) C_{g,2}^{(g,2)}(t)^*],\nonumber\\
\rho_{eg,ge} &=& \alpha_1^*\beta_1[|\beta_2|^2 C_{e,0}^{(e,0)}(t)^*+|\alpha_2|^2 C_{g,2}^{(g,2)}(t) C_{e,2}^{(e,2)}(t)^*],  \nonumber\\
\rho_{eg,eg} &=& |\beta_1|^2 [|\beta_2|^2 +|\alpha_2 C_{g,2}^{(g,2)}(t)|^2],\nonumber\\
\rho_{ef,ef} &=& |\alpha_2 \beta_1 C_{f,1}^{(g,2)}(t)|^2, \nonumber\\
\rho_{ee,ee} &=& |\alpha_2 \beta_1 C_{e,0}^{(g,2)}(t)|^2,
\end{eqnarray}
where $\rho_{ab,cd}(a,b,c,d=e,f,g)$ denotes $\rho_{|ab\rangle_{12},_{12}\langle cd|}$.
This density matrix $\rho$ has six non-zero eigenvalues:
\begin{eqnarray}\label{20}
  \lambda_1&=&\left|\alpha_2 \beta_1 C_{e0}^{(e0)}(t)\right|^2 ,\nonumber\\
\lambda_2&=&|\alpha_2\beta_1C_{f1}^{(g2)}(t)|^2,\nonumber\\
\lambda_3&=&|\alpha_1|^2 \left[|\beta_2 C_{f1}^{(e0)}(t)|^2+|\alpha_2C_{f3}^{(e2)}(t)|^2\right],\nonumber\\
\lambda_4&=&|\alpha_1|^2\left[|\beta_2  C_{g2}^{(e0)}(t)|^2+|\alpha_2 C_{g4}^{(e2)}(t)|^2\right],\nonumber\\
 \lambda_5&=&  \frac{1}{2}\left[|\alpha_1 \beta_2 C_{e0}^{(e0)}(t)|^2+|\alpha_1
   \alpha_2 C_{e2}^{(e2)}(t)|^2\right.\nonumber\\
   &&+\left.|\alpha_2 \beta_1
   C_{g2}^{(g2)}(t)|^2
   +|\beta_1 \beta_2|^2-\eta (t)\right],\nonumber\\
 \lambda_6&=& \frac{1}{2}\left[|\alpha_1 \beta_2 C_{e0}^{(e0)}(t)|^2+|\alpha_1
   \alpha_2 C_{e2}^{(e2)}(t)|^2\right.\nonumber\\
   &&+\left.|\alpha_2 \beta_1
   C_{g2}^{(g2)}(t)|^2
   +|\beta_1 \beta_2|^2+\eta (t)\right],
  \end{eqnarray}
where
  \begin{eqnarray}\label{21}
   \eta(t)&=& \left\{\left[|\alpha_1|^2 \left(|\beta_2
  C_{e0}^{(e0)}(t)|^2+|\alpha_2  C_{e2}^{(e2)}(t)|^2\right)\right.\right.\nonumber \\
   &&\left.-|\beta_1|^2 \left(|\alpha_2C_{g2}^{(g2)}(t)|^2+|\beta_2|^2\right)\right]^2 \nonumber \\
     & & + 4 |\alpha_1 \beta_1|^2
   \left(|\beta_2|^2 C_{e0}^{(e0)}(t)^*+|\alpha_2|^2
   C_{e2}^{(e2)}(t)^* C_{g2}^{(g2)}(t)\right)\nonumber\\
   &&\left.\times\left(|\beta_2|^2 C_{e0}^{(e0)}(t)+|\alpha_2|^2
   C_{e2}^{(e2)}(t) C_{g2}^{(g2)}(t)^*\right)\right\}^{1/2}.
  \end{eqnarray}
Finally, the entropies of two atomic system and field are given by
\begin{eqnarray}\label{22}
   S_f(t)=S_a(t)=-\mbox{Tr}[\rho_{12}\log(\rho_{12}(t))]=-\sum_{i=1}^6 \lambda_i \log(\lambda_i),
\end{eqnarray}
in the above formula, logarithms are taken to base two, as usual.

\begin{figure}
\centering
\includegraphics[width=6cm]{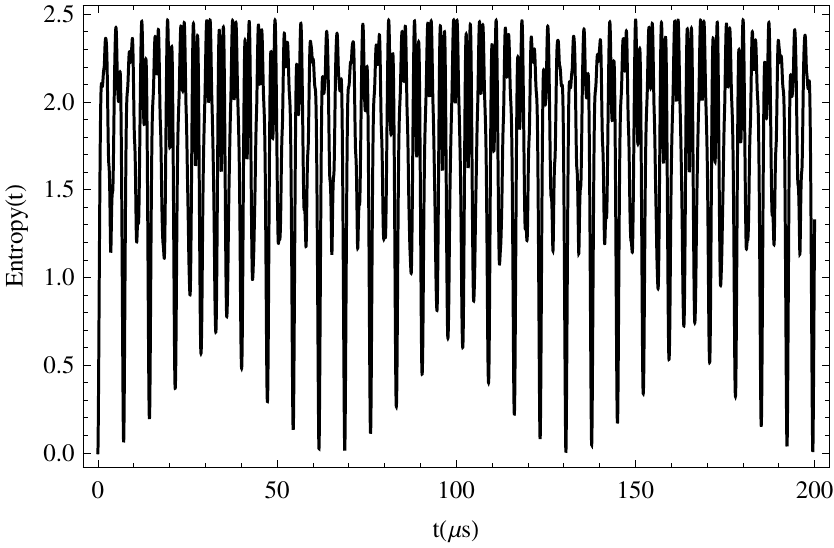}
\caption{Time evolution of the entropy of two three-level atoms interacting
with a single-mode of a cavity field with $\alpha_1 = \alpha_2 = \beta_1=\beta_2 = 1/\sqrt{2}$ and $g_1 = g_2 = g= 1 \mbox{MHz},\delta = 0$.} \label{entropy_delta0}
\end{figure}

\begin{figure}
\centering
\includegraphics[width=6cm]{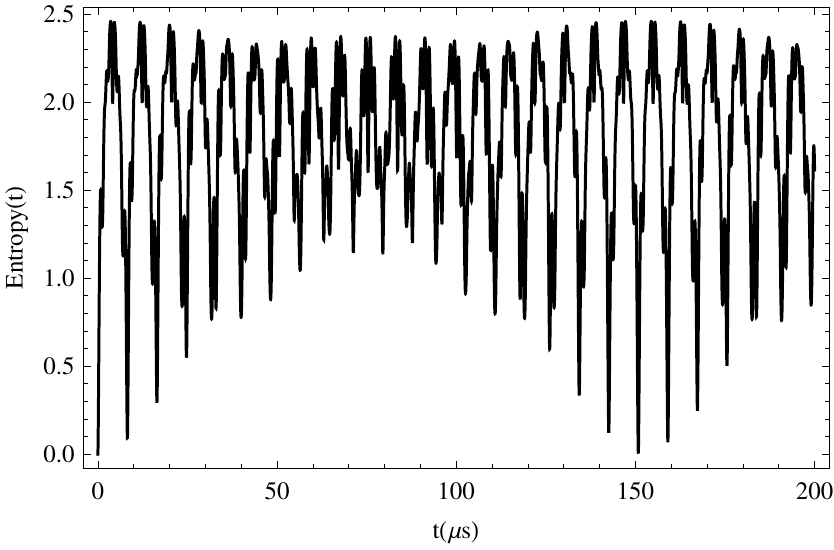}
\caption{Time evolution of the entropy of two three-level atoms interacting
with a single-mode of a cavity field with $\alpha_1 = \alpha_2 = \beta_1=\beta_2 = 1/\sqrt{2}$ and $g_1 = g_2 =g=  1 \mbox{MHz},\delta =3g$.} \label{entropy_delta3g}
\end{figure}

\section{Numerical results and discussion}
To reveal the properties of the entropy of the atomic and field systems, we numerically calculated the entropy according to Eqs.(20)-(22) with various values of the detuning $\delta$. The results are shown  in Figs. 2-8. These plots show the following interesting features. First, when $\delta$ is zero or small, the time evolution of the entropy oscillates with time $t$  but not explicitly shows periodicity. For larger $\delta$, the periodicity of the time evolution of the entropy is more and more obvious. This situation is similar to the case of a single three-level atom interacting with single cavity field [11]. Second, for large $\delta$, the periodicity showed by the time evolution of the entropy is multiple which means in a larger period the evolution of the entropy further periodically  shows variety modes. This is different from the case of a single three-level atom interacting with single-mode or two-mode cavity field [11,19] and can be attributed to entanglements of atom-atom, cavity-cavity and atom-cavity. Furthermore, even though we can not find the explicit formula of the period, we can approximately estimate from Figs. 2-8 for larger $\delta$, with the maximum periods of  $316 \mu s$, $630 \mu s$, $941 \mu s$ and $1257 \mu s$ corresponding to  $\delta=50g, 100g, 150g, 200g$, respectively. In addition, these data show the period linearly increased when  $\delta$ increased. Third, Figs. 2-8 also show that the maximum values of the entropy decreased when $\delta$ increased. This is natural because the larger detuning $\delta$ means that atomic system is weakly coupled to cavity fields, therefore, the degree of the entanglement is weakened.

\begin{figure}
\centering
\includegraphics[width=6cm]{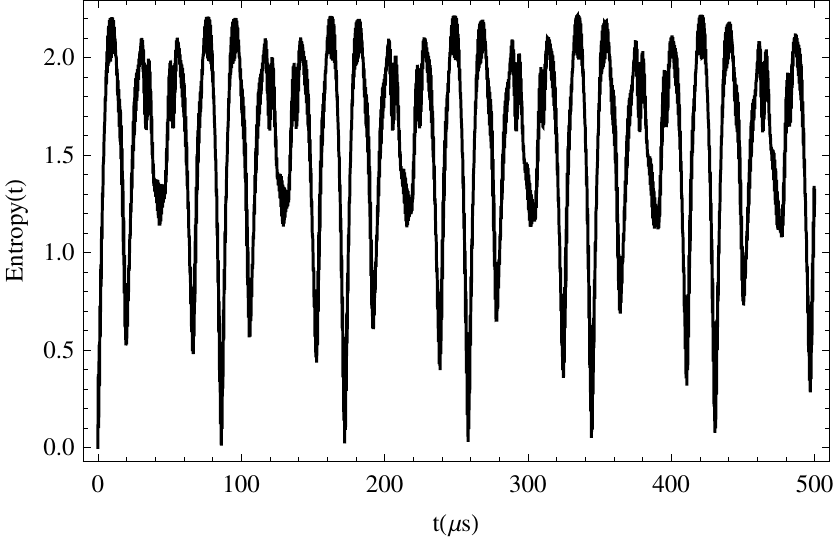}
\caption{Time evolution of the entropy of two three-level atoms interacting
with a single-mode of a cavity field with $\alpha_1 = \alpha_2 = \beta_1=\beta_2 = 1/\sqrt{2}$ and $g_1 = g_2 =g=  1 \mbox{MHz},\delta =10g$.} \label{entropy_delta10g}
\end{figure}

\begin{figure}
\centering
\includegraphics[width=6cm]{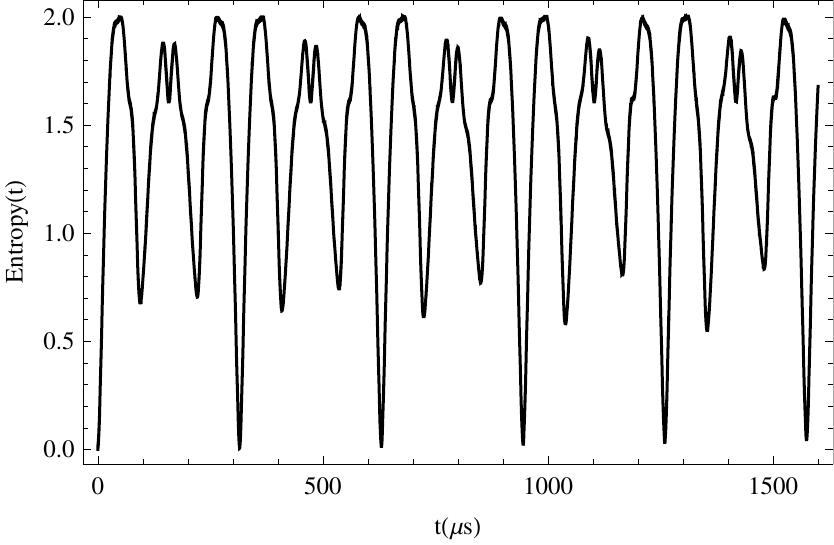}
\caption{Time evolution of the entropy of two three-level atoms interacting
with a single-mode of a cavity field with $\alpha_1 = \alpha_2 = \beta_1=\beta_2 = 1/\sqrt{2}$ and $g_1 = g_2 =g=  1 \mbox{MHz},\delta =50g$.} \label{entropy_delta50g}
\end{figure}

\begin{figure}
\centering
\includegraphics[width=6cm]{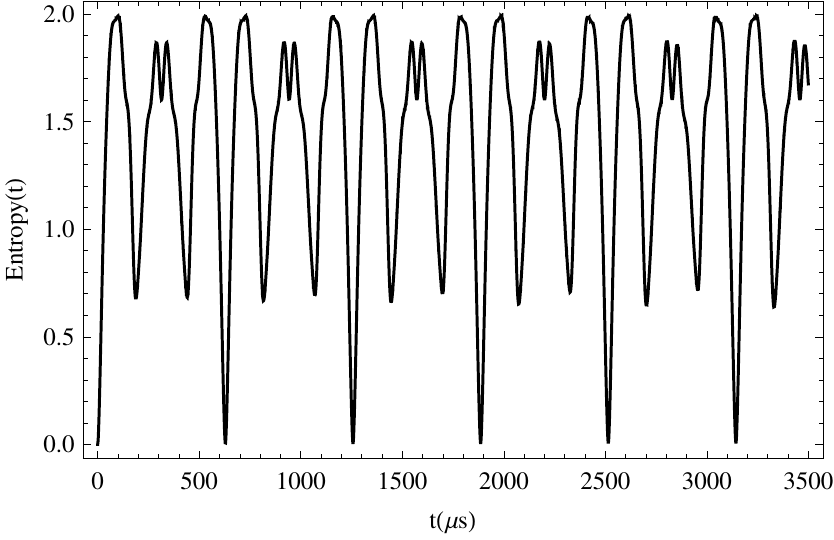}
\caption{Time evolution of the entropy of two three-level atoms interacting
with a single-mode of a cavity field with $\alpha_1 = \alpha_2 = \beta_1=\beta_2 = 1/\sqrt{2}$ and $g_1 = g_2 =g=  1 \mbox{MHz},\delta = 100g$.} \label{entropy_delta100g}
\end{figure}

\begin{figure}
\centering
\includegraphics[width=6cm]{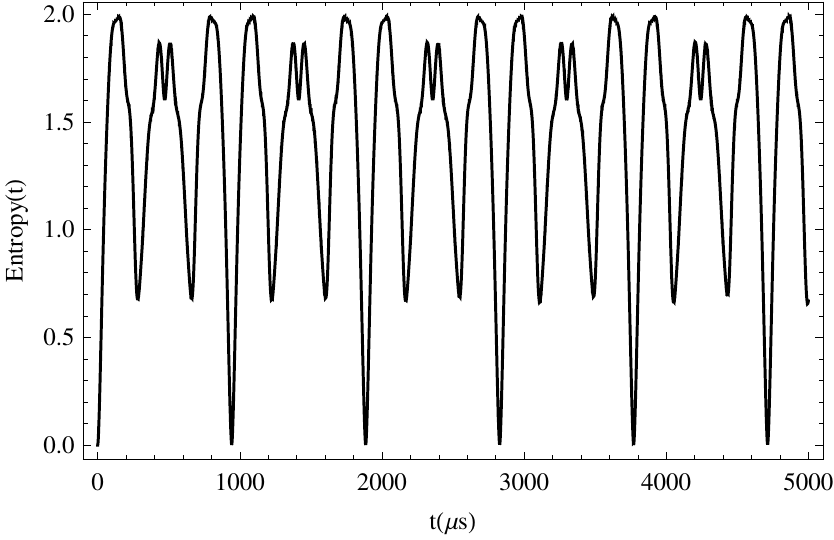}
\caption{Time evolution of the entropy of two three-level atoms interacting
with a single-mode of a cavity field with $\alpha_1 = \alpha_2 = \beta_1=\beta_2 = 1/\sqrt{2}$ and $g_1 = g_2 =g=1\mbox{MHz},\delta =150g$.} \label{entropy_delta150g}
\end{figure}

\begin{figure}
\centering
\includegraphics[width=6cm]{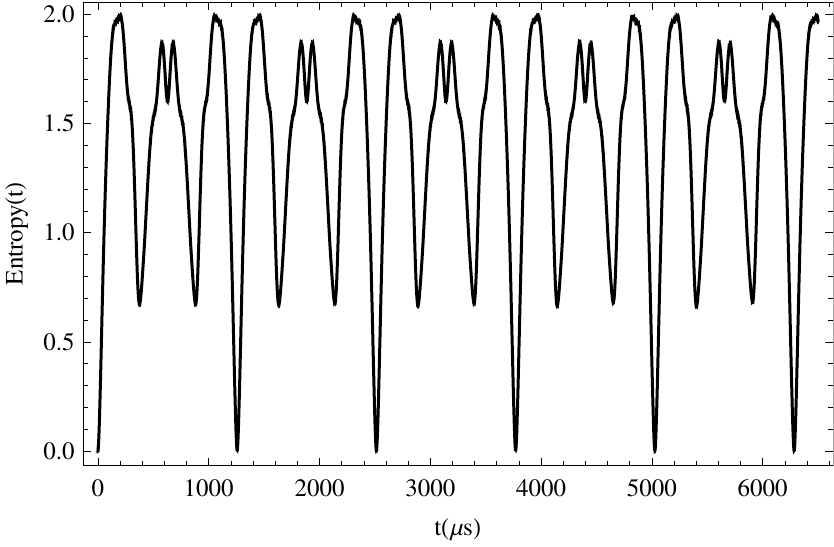}
\caption{Time evolution of the entropy of two three-level atoms interacting
with a single-mode of a cavity field with $\alpha_1 = \alpha_2 = \beta_1=\beta_2 = 1/\sqrt{2}$ and $g_1 = g_2 =g=  1 \mbox{MHz},\delta =200g$.} \label{entropy_delta200g}
\end{figure}

\section{Concluding remarks}
Using FMHA and two-photon Jaynes-Cummings model, we have found the analytical expression of the reduced density matrix for two entangled atoms after an entanglement swapping with two cavity fields. The explicit formula of the entropy for atomic and field systems are also obtained. It is a distinct feature that the entropy evolution of atomic system and fields exhibits multi-periodicity. Furthermore, maximum  period linearly increased as  $\delta$ increased for large detuning $\delta$.

\vspace{6mm} {\bf Acknowledgments}: This work is supported
 by Natural Science Basic Research Plan in the Shaanxi
Province of China (program no: SJ08A13), the Natural Science
Foundation of the Education Bureau of Shaanxi Province, China
under Grant O9jk534. WBC thanks the CAPES, Brazilian agency, for the partial support.\\


\begin{thebibliography}{99}
\bibitem{1}  M. A. Nielsen and I. L. Chuang 2000 Quantum Computation and
Quantum Information (Cambridge: Cambridge University Press).
\bibitem{2} B. Buck, C.V. Sukumar, Phys. Lett. A 81 (1981) 132.
\bibitem{3} C. V. Sukumar, B. Buck, Phys. Lett. A 83 (1981) 211.
\bibitem{4} S. J. D. Phoenix, P. L. Knight, Phys. Rev. A 44 (1991) 6023.
\bibitem{5} S. J. D. Phoenix, P. L. Knight, Phys. Rev. Lett. 66 (1991) 2833.
\bibitem{6} J. Gea-Banacloche, Phys. Rev. Lett. 65 (1990) 3385.
\bibitem{7} J. Gea-Banacloche, Phys. Rev. A 44 (1990) 5913.
\bibitem{8} A. J. Wonderen, F. Farhadmotamed, K. Lendi, J. Phys. A: Math. Gen. 31 (1998)
3395.
\bibitem{9} M. F. Fang, H. E. Liu, Phys. Lett. A 200 (1995) 250.
\bibitem{10} Xiang Liu, Physica A 286 (2000) 588.
\bibitem{11} C. Huang,  L. Tang, F. Kong, J. Fang, M. Zhou, Physica A 368 (2006) 25.
\bibitem{12} M. Abdel-Aty, J. Phys. B 33 (2000) 2665.
\bibitem{13} M. Abdel-Aty, A.-S.F. Obada, Eur. Phys. J. D 23 (2000) 155.
\bibitem{14} M. A. Can, O. Cakir, A. Klyachko, A. Shumovsky, J. Opt. B 6 (2004) S13.
\bibitem{15} A.-S. F. Obada, A. A. Eied, G. M. Abd Al-Kader, J. Phys. B 41 (2008) 195503.
\bibitem{16} A.-S. F. Obada, A. A. Eied, Opt.Commun. 282 (2009) 2184.
\bibitem{17} A. D. dSouza, W. B. Cardoso, A. T. Avelar and B. Baseia, Phys. Scr. 80 (2009) 065009.
\bibitem{18} A. H. Toor and M. S. Zubairy, Phys. Rev. A 45 (1992) 4951.
\bibitem{19} A. D. dSouza, W. B. Cardoso, A. T. Avelar, B. Baseia, Physica A 388 (2009) 1331.
\bibitem{20} H. Araki, E. Lieb, Commun. Math. Phys. 18 (1970) 160.
\end{thebibliography}
\end{document}